\documentclass[preprintnumbers,amsmath,amssymb,floatfix,11pt,prd,superscriptaddress,nofootinbib]{revtex4}
\usepackage{graphicx}
\usepackage{epsfig}
\usepackage{bm}
\usepackage{amsfonts}

\def\ben{\begin{equation}}
\def\een{\end{equation}}

 \def\bd{\begin{document}} \def\ed{\end{document}}
\def\ds{\documentstyle} \let\fr=\frac \let\bl=\bigl \let\br=\bigr
\let\Br=\Bigr \let\Bl=\Bigl
\let\bm=\bibitem
\let\na=\nabla
\let\pa=\partial \let\ov=\overline
\newcommand{\be}{\begin{equation}}
\newcommand{\ee}{\end{equation}}
\def\ba{\begin{array}}
\def\ea{\end{array}}
\def\ft#1#2{{\textstyle{\frac{\scriptstyle #1}{\scriptstyle #2} } }}
\def\fft#1#2{{\frac{#1}{#2}}}
\def\del{\partial}
\def\vp{\varphi}
\def\sst#1{{\scriptscriptstyle #1}}
\def\oneone{\rlap 1\mkern4mu{\rm l}}
\def\td{\tilde}
\def\wtd{\widetilde}
\def\ie{{\it i.e.\ }}
\def\dalemb#1#2{{\vbox{\hrule height .#2pt
        \hbox{\vrule width.#2pt height#1pt \kern#1pt
                \vrule width.#2pt}
        \hrule height.#2pt}}}
\def\square{\mathord{\dalemb{6.8}{7}\hbox{\hskip1pt}}}
\newcommand{\ho}[1]{$\, ^{#1}$}
\newcommand{\hoch}[1]{$\, ^{#1}$}
\newcommand{\bea}{\setlength\arraycolsep{2pt} \begin{eqnarray}}
\newcommand{\eea}{\end{eqnarray}}
\newcommand{\ra}{\rightarrow}
\newcommand{\lra}{\longrightarrow}
\newcommand{\Lra}{\Leftrightarrow}
\newcommand{\bp}{\tilde \beta^\prime}
\newcommand{\tr}{{\rm tr} }
\newcommand{\Tr}{{\rm Tr} }
\def\0{{\sst{(0)}}}
\def\1{{\sst{(1)}}}
\def\2{{\sst{(2)}}}
\def\3{{\sst{(3)}}}
\def\4{{\sst{(4)}}}
\def\5{{\sst{(5)}}}
\def\6{{\sst{(6)}}}
\def\7{{\sst{(7)}}}
\def\8{{\sst{(8)}}}
\def\m{{\sst{(m)}}}
\def\n{{\sst{(n)}}}
\def\cA{{{\cal A}}}
\def\cB{{{\cal B}}}
\def\cF{{{\cal F}}}
\def\cG{{{\cal G}}}
\def\cH{{{\cal H}}}
\def\tV{\widetilde V}
\def\tW{\widetilde W}
\def\tH{\widetilde H}
\def\tE{\widetilde E}
\def\tF{\widetilde F}
\def\tA{\widetilde A}
\def\im{{{\rm i}}}
\def\tY{{{\wtd Y}}}
\def\ep{{\epsilon}}
\def\vep{{\varepsilon}}
\def\bD{{{\bar D}}}
\def\R{{{\mathbb R}}}
\def\C{{{\mathbb C}}}
\def\H{{{\mathbb H}}}
\def\CP{{{\mathbb C}{\mathbb P}}}
\def\RP{{{\mathbb R}{\mathbb P}}}
\def\Z{{{\mathbb Z}}}
\def\bA{{{\mathbb A}}}
\def\bB{{{\mathbb B}}}
\def\bC{{{\mathbb C}}}
\def\bD{{{\mathbb D}}}
\def\bE{{{\mathbb E}}}
\def\bZ{{{\mathbb Z}}}
\def\Re{{{\frak{Re}}}}
\def\Im{{{\frak{Im}}}}
\def\cosec{{\,\hbox{cosec}\,}}
\def\Gm{{\Gamma_{\!\! -}}}
\def\Gp{{\Gamma_{\!\! +}}}
\def\stan{{standard }}
\def\nonstan{{supernumerary }}
\def\p{{\partial}}
\def\kdel#1{{\fft{\del}{\del#1}}}

\def\bog{{Bogomolny }}
\def\om{{\omega}}
\newcommand{\un}{\underline}
\def\R{\hbox{{\rm I}\kern-0.2em{\rm R}\kern0.2em}}
\def\D{\hbox{{\rm I}\kern-0.2em{\rm D}\kern0.2em}}
\def\a{\alpha} \def\o{\omega} \def\w{\wedge}
\def\b{\beta}         \def\rf{\rfloor}
\def\e{{\rm e}}
\def\ld{\lambda} \def\Ld{\Omega}
\def\d{{\rm d}}
\def\dsy{\displaystyle}
\def\de{\delta}
\def\ep{\epsilon}
\def\g{\gamma}
\def\be{\begin{equation}}
\def\ee{\end{equation}}
\def\X{{\cal X}} \def\U{{\cal U}}
\def\p{\partial}
\def\({\left(}
\def\){\right)}
\def\[{\left[}
\def\]{\right]}
\def\bc{\begin{center}}
\def\ec{\end{center}}
\def\EL{Euler-Lagrange}
\def\ph{\phantom}
\def\wh{\widehat}
\def\n{\noindent}
\newcommand{\nnr}{\nonumber \\}
\newcommand{\pd}{\partial}
\newcommand{\ud}{\textrm{d}}
\newcommand{\dTH}{T^{\prime \, 0}_\textrm{H}}
\newcommand{\dOi}{\Omega^{\prime \, 0}_i}
\newcommand{\bx}{{\bf x}}
\begin{document}

\title{Quantized Black Hole and Heun function }

\author{\textbf{D. Momeni}}
\email{d.momeni@yahoo.com} 
\affiliation{Eurasian International
Center for Theoretical Physics,Eurasian National University, Astana
010008, Kazakhstan}
\author{\textbf{Koblandy Yerzhanov}}
\affiliation{Eurasian International Center for Theoretical Physics,Eurasian National University, Astana 010008,
Kazakhstan}
\author{\textbf{Ratbay Myrzakulov}}
\email{rmyrzakulov@csufresno.edu;
rmyrzakulov@gmail.com}\affiliation{Eurasian International Center for Theoretical Physics,Eurasian National University, Astana 010008,
Kazakhstan}
\begin{abstract}
\vspace*{1.5cm} \centerline{\bf Abstract} \vspace*{6mm}
 Following the simple proposal by He and Ma for quantization of a black hole(BH) by Bohr's idea about the atoms,we discussed the solvability of the wave equation for such a BH. We superficial solved the associated Schrodinger equation. The eigenfunction problem reduces to  HeunB $H$ differential
 equation which is a natural generalization of the hypergeometric
differential equation. In other words, the spectrum can be determined by
solving the Heun's differential equation.
\end{abstract} \vspace{6mm}

\maketitle

\newpage

\section{Introduction}

In 2010,Verlinde inspired directly from the Holography
concept\cite{Hooft,Susskind} ,proposed an idea \cite{Verlinde} about the
quantum origin of the gravity and the second law of inertia. In this scenario, the key objects are the concept of the change in entropy stored on the holographic screen, when the test particle moves toward the screen and the entropic force comes from the polymer physics. In the simplest form, the entropic force try to recover the Newton's gravitational theory and it has been shown that why we can believe to gravity as an thermodynamical quantity. There are many papers in literatures about this idea and it's possible extensions \cite{entropic}. The simple mathematics behind entropic scenarion excited many disputations about the validity of it's
 arguments. Essentially the relation between thermodynamics and the black hole physics is an old idea\cite{bekenstein,hawking}
 and indeed there is a deep relation between general relativity (GR) and thermodynamical structure of the field
eqs\cite{jacobson,paddy}. It has been proved that the Einstein field equations avaluated on the horizon can be interpreted as the first law of the thermodynamics for BH.  Discovering of the Hawking
 radiation and the famous area-entropy formula for BHs and it's extensions to the power law and logarithmic corrected entropy function opened new windows to   the new abilities to demeaning the BH horizon as a thermodynamical system
 sustained this  conjecture. If we accept  the
 Velinde idea ,a test particle which obeys from the uncertainly
 principle near to a holographic  screen (which is an
 equipotential surface) changes the entropy of the screen slightly
 and as which is argued original by Verlinde this amount of the entropy which is proportional
 to the displacement,  causes a force which
 is responsible for the attractive force between the test
 particle and the screen. Some problems as the locality of the energy
 and the temperature remain unsolved. For example it is difficult to interpret in any arbitrary case the Unruh temperature of the accelerated observer exactly in the same manner as the horizon temperature $T_{BH}$.  Some authors showed that the  simple form of the gravitational force in the verlinde idea is not unique, and by taking some
 additional terms which each term causes some corrections to the BH
 entropy formula, for example the log corrections or volume
 corrections \cite{Modesto}. Such kinds of the entropy corrections  come  from the LQG \cite{LQG},and the
 similarity of the new force terms to the MOND approach\cite{MOND},all of them show
 us that (may be) we work in an  equivalent formulation. The quantum
 corrections to the Newton's gravitation law and the unfamiliar and
 strange uncertainty principle arisen from it,all are 
 lugubrious. Anyway ,from many papers on this topics one which
 acclaimed about the quantization of the BH, attracted us. In \cite{HeMa}
 the authors under simple conditions and  from the Bohr's
 quantization, construct some expression for the energy, the surface
 area, the Compton's wavelength, the surface gravity, quantized
 temperature. We do not repeat their results and referring the reader
 to the \cite{HeMa}. In this work we want to calculate the 
 wave function for this quantized energy spectrum of the BH area. We
 assume that this energy spectrum (which possesses also low energy continuum
 behavior),corresponds to a particle with mass of order  Planck mass and it's
radial potential varies by  a power of $\frac{2}{3}$. The wave function will be
 calculated in terms of the more complicated functions as the usual
quantum mechanical radial  wave functions.

 The plan of this paper is as follow. In section II we obtain the potential function using WKB method. In section III we solve the wave equation using the  series method. In section IV we apply the Poincare asymptotic method to recover the exact wave function. In section V we calculate the expectation values of the kinetic and potential energies using Hellmann–Feynman theorem. We summarize and conclude in last section.
\section{The general potential for a test particle with energy spectrum $E_{n}=M_{p}c^{2}\sqrt{n}$}

As it was stated by He-Ma,the general energy spectrum for a 
quantized BH is
relation\cite{HeMa}:
\begin{eqnarray}
\begin{aligned}
E_{n}=M_{p}c^{2}\sqrt{n}.
\end{aligned}
\end{eqnarray}
We know that the Bohr scenario for quantization of a Hydrogen atom
is valid only in large quantum numbers,namely it is the asymptotic
limit of the general WKB approximation. As a
simple consequence of imposing the Bohr-Sommerfeld quantization to a
general potential function both one dimensional or completely
spherically symmetric ones,we know that this energy spectrum can be produced
from the following potential function

\begin{eqnarray}
\begin{aligned}
V=V_{0}|\overrightarrow{r}|^{p}\Rightarrow E_{n}\approx
n^{\frac{2p}{p+2}}
\end{aligned}
\end{eqnarray}

 reminiscence the validity of (1) in the large quantum numbers $n$
 and comparing to the (2),we deduce that the general potential term
 must be
\begin{eqnarray}
\begin{aligned}
V=V_{0}|\overrightarrow{r}|^{\frac{2}{3}}.
\end{aligned}
\end{eqnarray}
The potential function (3) is not bound. In Figure 1 we plot the potential function. It shows that in the semi classical limit if the BH want to remains stable, firstly the matter waves must be standing waves and no bound states is needed for such a quantized BH.

\begin{figure}
\centering
 \includegraphics[scale=0.3] {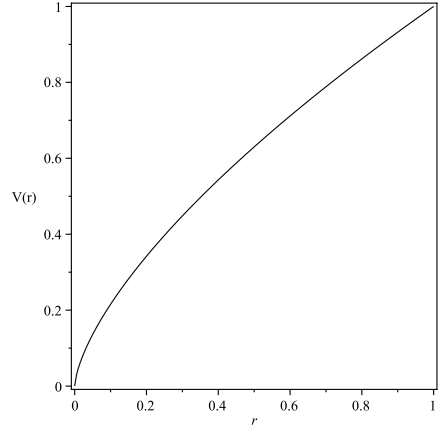}
  \caption{The potential function $V(r)$ for a particle in the quantized BH.  }
  \label{1.eps}
\end{figure}

\section{ Schr\"{o}dinger wave equation }

In this section we will solve the wave equation for a quantized BH. We assume that the BH behaves like a paricle with mass $M$ ,and write the static wave functions
$\phi(r,\theta,\phi)=\frac{u(r)}{r}Y_{lm}(\theta,\phi)$,in usual
spherically symmetric coordinates and a  normal separation of the
variables,we obtain the radial equation

\begin{eqnarray}
\begin{aligned}
-\frac{\hbar^{2}}{2M}
u''+(\frac{\hbar^{2}l(l+1)}{2Mr^{2}}+V_{0}r^{\frac{2}{3}}-M_pc^{2}\sqrt{n})u=0.
\end{aligned}
\end{eqnarray}
It is adequate to introducing a new set of variable   $x=\alpha r$
where $\alpha=(\frac{2MV_{0}}{\hbar^{2}})^{\frac{3}{8}}$ and a new
parameter $b=M_p c^2(\frac{2M}{\hbar^2 V_0^3})^{\frac{1}{4}}$. In
these variables we obtain
\begin{eqnarray}
\begin{aligned}
\frac{d^2 u}{d
x^2}+(-\frac{l(l+1)}{x^{2}}-x^{\frac{2}{3}}+b\sqrt{n})u=0.
\end{aligned}
\end{eqnarray}

The general solution for this diffrential equation,is in the form of
HeunB functions\cite , denoted by $H$ \cite{Heun} multiplied by some functions
\begin{eqnarray}
\begin{aligned}
u(x)=e^{-f_n}\Big[c_{1}x^{-l}H(-\frac{3}{2}-3l,b\sqrt{\frac{3n}{2}},\frac{3}{8}nb^2,0,-\frac{3}{2}x^{2/3})
+c_{2}x^{l+1}H(\frac{3}{2}+3l,b\sqrt{\frac{3n}{2}},\frac{3}{8}nb^2,0,-\frac{3}{2}x^{2/3})\Big]
\end{aligned}
\end{eqnarray}
where
\begin{eqnarray}
\begin{aligned}
f_n=-\frac{3}{4}x^{\frac{3}{2}}(x^{\frac{3}{2}}-b\sqrt{n}).
\end{aligned}
\end{eqnarray}
In Figure 2, we plot the function $f_n$ as a function of $x$ with $b=1$ for diffrent values of n.
\begin{figure}
\centering
 \includegraphics[scale=0.3] {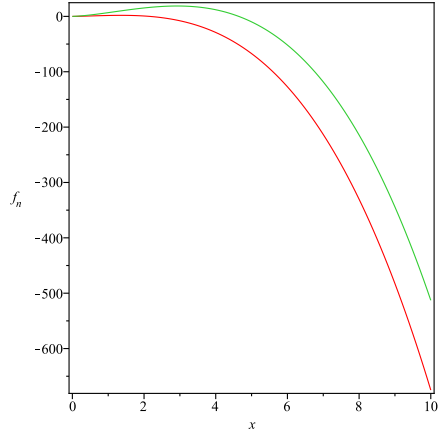}
  \caption{The $f_n$ function  for $n=10(Red),100(Green)$.  }
  \label{2.eps}
\end{figure}

 We can set $c_{2}=0$ for square integrability
of the wave function. Thus the total wave function is:
\begin{eqnarray}
\begin{aligned}
\phi(x,\theta,\phi)=Ne^{-f_n}x^{-l-1}H(\pm (\frac{3}{2}+3l)
,b\sqrt{\frac{3n}{2}},\frac{3}{8}nb^2,0,-\frac{3}{2}x^{2/3})Y_{lm}(\theta,\phi)
\end{aligned}
\end{eqnarray}
In Figure 3, we plot the radial wave function $u(x)$ for some values of the parameters.
\begin{figure}
\centering
 \includegraphics[scale=0.3] {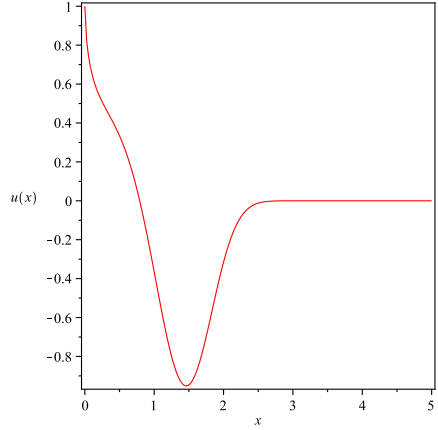}
  \caption{The normalizable radial function  $u(x)$ for s-wave $l=0,n=10$ , $b=\sqrt{\frac{2}{3}}$.  }
  \label{3.eps}
\end{figure}

\section{Poincare's Asymptotic method}

 We write out the local expansion of the coefficients of Eq.
(5) near a regular singularity $x=\infty$:
\begin{eqnarray}
\begin{aligned}
u^{\infty}(x)=Ax^{1/2}Z_{\frac{3}{8}}\Big(\frac{3}{4}x^{\frac{4}{3}}\Big),
\end{aligned}
\end{eqnarray}
Where the $Z_{\nu}(ax)$ is apiece
 of the Bessel functions of the second
kind.  Also we  introduce a pair of Frobenius
solutions in a neighborhood of an apparent singularity $x=0$ that is
fixed by the following behavior:
\begin{eqnarray}
\begin{aligned}
u_{0}(x)=c_{1}x^{-l}+c_{2}x^{l+1}.
\end{aligned}
\end{eqnarray}
Note that apparent singular points $x=0,\infty$ play an important role
in  general monodromy properties of linear Fuchsian
systems.\\
Now we suppose that the general solution for (5) is
\begin{eqnarray}
\begin{aligned}
u(x)=Ax^{\frac{3}{2}+l}Z_{\frac{3}{8}}\Big(\frac{3}{4}x^{\frac{4}{3}}\Big)y(x).
\end{aligned}
\end{eqnarray}
Remember to mind that in the asymptotic regime $x=\infty$ we have
\begin{eqnarray}
\begin{aligned}
Z_{\frac{3}{8}}\Big(ax^{\frac{4}{3}}\Big)\approx
\sqrt{\frac{\pi}{2a}}\frac{e^{-ax^{\frac{4}{3}}}}{x^{\frac{2}{3}}}+O\Big(\frac{1}{x^2}\Big).
\end{aligned}
\end{eqnarray}
Substituing (11) in (5) we obtain the following diffrential equation for unknown
function $y(x)$:

\begin{eqnarray}
\begin{aligned}
 x^{2}y''+
\left( 2\,lx+\frac{5}{3}\,x-2\,{x}^{\frac{7}{3}} \right) y' + \left( -\frac{1}{3}\,l-{\frac
{5}{36}}-2l\,{x}^{\frac{4}{3}}-2\,{x}^{\frac{4}{3}}+b\sqrt {n}{ x}^{2} \right) y
 =0.
\end{aligned}
\end{eqnarray}
It is very phenomenal that we anew envisage the Heun solution
completely in accordance
 with the previous direct method :
\begin{eqnarray}
\begin{aligned}
y(x)={{\rm e}^{\frac{3}{4}\,{x}^{\frac{2n}{3}}b}}\Big[c_{{1}}\sqrt [6]{x}{\it
H} \left( \frac{3}{2}+3\,l ,\frac{1}{2}\,b\sqrt {6n},\frac{3}{8}\,n{b}^{2},0,-\frac{1}{2}\,\sqrt {6}{x}^{\frac{2}{3}}
 \right)\\\nonumber + c_{{2}}{x}^{
-\frac{5}{6}-2\,l}{\it H} \left( -\frac{3}{2}-3\,l,\frac{1}{2}\,b\sqrt
{6n},\frac{3}{8}\,n {b}^{2},0,-\frac{1}{2}\,\sqrt {6}{x}^{\frac{2}{3}} \right)\Big].
\end{aligned}
\end{eqnarray}

\section{Expectation Value of the kinetic energy for $l=0$}

The normalization constant $N$ in (7) is determined from the
requirement that:
\begin{eqnarray}
\begin{aligned}
\int_{0}^{\infty}|u(x)|^{2}dx=1.
\end{aligned}
\end{eqnarray}
Some useful expectation values $<r>,<r^{2}>,..$ and the virial
theorem for the  potential  can be obtained by applying
Hellmann–Feynman theorem (HFT). Assuming that the Hamiltonian
$\mathcal{H}$ for a particular quantum mechanical system is a
function of some parameter q, let $E(q)$ and $\phi(q)$ be the
eigenvalues and the eigenfunctions of the Hamiltonian $\mathcal{H}$.
Then, the Hellmann–Feynman theorem states that\cite{HF}
\begin{eqnarray}
\begin{aligned}
\frac{\partial E(q)}{\partial q}  =<\phi(q)\mid\frac{\partial
\mathcal{H}}{\partial q}\mid\phi(q)>.
\end{aligned}
\end{eqnarray}
The effective Hamiltonian for the Quantized BH's wave function in
case $l=0$ is

\begin{eqnarray}
\begin{aligned}
\mathcal{H}=-\frac{\hbar^{2}}{2M}
\frac{d^2}{dr^2}+V_{0}r^{\frac{2}{3}}.
\end{aligned}
\end{eqnarray}
we let $q=M_{p}$   in (15) , then,
\begin{eqnarray}
\begin{aligned}
\frac{\partial E(M_{p})}{\partial M_{p}}
=<\phi(M_{p})\mid\frac{\partial \mathcal{H}}{\partial
M_{p}}\mid\phi(M_{p})>,
\end{aligned}
\end{eqnarray}
while,(Only valid in the semi-Classical limit $n>>1$)
\begin{eqnarray}
\begin{aligned}
\frac{\partial E(M_{p})}{\partial M_{p}}=c^{2}\sqrt{n}
\end{aligned}
\end{eqnarray}
Therefore, the expectation values for $
\mathcal{T}=-\frac{\hbar^{2}}{2M_{p}} \frac{d^2}{dr^2}$(using HFT)
in this case is:
\begin{eqnarray}
\begin{aligned}
<\mathcal{T}>=\frac{1}{3}<V_{0}r^{\frac{2}{3}}>\approx
M_{p}c^{2}\sqrt{n}.
\end{aligned}
\end{eqnarray}
Completely in accordance with the results of \cite{Popov}. we can also deduce
the expectation value of the momentum-square and the potential
energy in this limit.
\section{Conclusion}
Quantization of a black hole is a hot topic in contemporary physics. Recently based on the idea of Bohr and extension of the Verlinde entopic idea to black hole with spherical horizon as the holographic screen, a simple model for quantization of such static black holes presented. In this scenario, by applying the ansatz of the Verlinde about the change of the total entropy of a holographic screen via the formula $\Delta S=\frac{4\pi k_B \delta R}{\lambda_c}$, where the displacement is of order of the horizon size of the black hole and by identifying a Compton wavelength to the black hole of order $\lambda_c=\frac{\hbar}{M}$, it is possible to obtain the energy spectrum of the quantized black hole with the formula $E_n=M_p c^2 \sqrt{n}$. In this work we presented the possible form of the potential for this eigenvalue of the Hamiltonian of the quantized system as a simple quantum mechanical system. We wrote the Shrodinger wave equation for a particle with mass M  and associate to it a potential function of the form $V(r)\propto r^{\frac{2}{3}}$. We shown that it is possible to solve the wave equation appropriately by series method. We obtained the exact solution in terms of the HeunB
function via direct solving the ODE and also using the Poincare's
asymptotic method.  We
investigated the wave function's behavior and discussed some properties of this wave function in terms of the Heun functions. Further  the expectation value of the
kinetic and potential energies computed. Our work shown the relation b

\end{document}